\documentclass[onecolumn]{aa}
\usepackage{graphicx}
\begin{document}
   \title{M87 as a misaligned Synchrotron-Proton Blazar}

   \author{A. Reimer
          \inst{1}
          \and
          R.J. Protheroe\inst{2}
          \and
          A.-C. Donea\inst{2}
          }

   \offprints{A. Reimer}

   \institute{Institut f\"ur Theoretische Physik, Lehrstuhl IV:
              Weltraum- \& Astrophysik, Ruhr-Universit\"at Bochum,
              D-44780 Bochum, Germany\\
              \email{afm@tp4.rub.de}
         \and
              Department of Physics and Mathematical Physics,
              The University of Adelaide, Adelaide, SA 5005, Australia\\
             \email{rprother@physics.adelaide.edu.au, adonea@physics.adelaide.edu.au}
             }

   \date{Received August 27, 2003; accepted February 3, 2004}

   \abstract{ The giant radio galaxy M87 is usually classified as
   a Fanaroff-Riley class I source, suggesting that M87 is a
   mis-aligned BL Lac object. Its unresolved nuclear region emits
   strong non-thermal emission from radio to X-rays which has
   been interpreted as synchrotron radiation.
   In an earlier
   paper we predicted M87 as a source of detectable gamma ray
   emission in the context of the hadronic Synchrotron-Proton
   Blazar (SPB) model.
   The subsequent tentative detection of TeV
   energy photons by the HEGRA-telescope array would, if
   confirmed, make it the first radio galaxy to be detected at
   TeV-energies.  We discuss the emission from the unresolved
   nuclear region of M87 in the context of the SPB model, and
   give examples of possible model representations of its
   non-simultaneous spectral energy distribution.  The low-energy
   component can be explained as synchrotron radiation by a
   primary relativistic electron population that is injected
   together with energetic protons into a highly magnetized
   emission region.  We find that the $\gamma$-ray power output
   is dominated either by $\mu^\pm$/$\pi^\pm$ synchrotron or
   proton synchrotron radiation depending on whether the primary
   electron synchrotron component peaks at low or high energies,
   respectively.  The predicted $\gamma$-ray luminosity peaks at
   $\sim$100 GeV at a level comparable to that of the low-energy
   hump, and this makes M87 a promising candidate source for the
   newly-commissioned high-sensitivity low-threshold Cherenkov
   telescopes H.E.S.S., VERITAS, MAGIC and CANGAROO III. Because
   of its proximity, the high-energy spectrum of M87 is
   unaffected by absorption in the cosmic infrared (IR)
   background radiation field, and could therefore serve as a
   template spectrum for the corresponding class of blazar if
   corrected for mis-alignment effects. This could significantly
   push efforts to constrain the cosmic IR radiation field
   through observation of more distant TeV-blazars, and could
   have a strong impact on blazar emission models.  If M87 is a
   mis-aligned BL-Lac object and produces TeV-photons as recently
   detected by the HEGRA-array, in the context of the SPB model
   it must also be an efficient proton accelerator.
    }

   \maketitle
%

\section{Introduction}

The Fanaroff-Riley (FR) class I giant radio galaxy
M87, situated nearly at the center of the Virgo cluster,
was the first extragalactic jet to be discovered (\cite{old}),
and has since then been intensively observed at all wavelengths.
Its proximity ($\sim 16.3$ Mpc; \cite{Cohen2000}) makes it an interesting laboratory
for testing and understanding extragalactic jets of radio-loud Active Galactic Nuclei (AGN)
and their powering engines.

Because M87 is sufficiently near for ultra-high-energy cosmic
rays (UHECRs) to be little affected by the GZK-cutoff at $\sim
5\times 10^{19}$ eV (\cite{Gre66}, \cite{Zat66}), and because its
size scales could allow magnetic confinement of the most
energetic cosmic rays (\cite{Hillas}), M87 has long been
considered as one of the prime candidate sources of high energy
cosmic rays. This idea has recently received some support by the
suggestion that possible clustering observed in the arrival
directions of the UHECRs can be understood in terms of deflection
of UHECRs from M87 by our Galaxy's magnetized wind assuming a
Parker-spiral magnetic structure (\cite{Ahn2000};
\cite{Biermanetal2001}). It appears, however, that such
deflection in the Galactic wind may be insensitive to the
direction of the cosmic ray sources (\cite{Billoir2000}).
Nevertheless, \cite{Ray2003} found
by using the cosmic ray output predicted in the SPB model
 that M87 could explain the observed
UHECR flux if the magnetic field topology between M87 and our
Galaxy were favourable.  They found that UHECR with energies
above $10^{20}$ eV could easily be produced because neutrons
produced in the pion photoproduction process would be
relativistically beamed along the jet direction and Doppler
boosted in energy.  Even though M87's jet is mis-aligned with
respect to our line-of-sight, these Doppler boosted neutrons
escape from the jet and decay into UHECRs which maintain their
Doppler boosted energies and may propagate in all directions,
including towards our Galaxy if the magnetic field topology were
favourable.
Of course we note that M87's nuclear region is not
the only possible source of the UHECRs observed at Earth; see
e.g. \cite{ProtheroeClay2004} for a recent review.

According to the unification model of AGN
(e.g. \cite{UrryPadovani95}) FR~I radio galaxies, with their jet
axis at a large angle to our line-of-sight, are the parent
population of BL Lac objects whose jets are closely aligned to
our line-of-sight. This motivates us to consider M87's nuclear
region as a mis-aligned blazar of BL Lac type.  The spectral
energy distribution (SED) of BL Lac objects can usually be
explained satisfactorily by either leptonic or hadronic blazar
emission models.
\cite{BaiLee2001} have
discussed M87 on the basis of the leptonic Synchrotron-Self
Compton (SSC) model where synchrotron photons produced by
interactions of relativistic electrons with the ambient magnetic
field serve as the target photons for inverse Compton scattering
by the same electrons. By interpreting the non-thermal radiation
from the radio to the X-ray band as synchrotron emission with
luminosity peaking in the far-ultraviolet, the authors considered
M87 to be a mis-aligned high-frequency peaked BL Lac (HBL), and
predicted $\gamma$-ray emission with an inverse Compton peak at
$\sim$100 GeV. The predicted inverse Compton flux is consistent
with the recent HEGRA detection of M87 (\cite{HEGRAdet}, see
Sect.~2).  Detectable TeV-emission from Comptonization of
galactic photon fields has recently been suggested by
\cite{Stawarz2003}.  In contrast to former models, they consider,
however, the large scale jet to be the site of $\gamma$-ray
production.

While in leptonic models a relativistic electron-positron plasma is
usually assumed to be responsible for the non-thermal jet
radiation, in hadronic models a relativistic proton-electron (p
e$^-$) plasma is assumed to be the main constituent of the jet
material.  In the hadronic Synchrotron-Proton Blazar (SPB) model,
proposed recently by, e.g., \cite{MP2001}, accelerated protons
interact with the synchrotron radiation field produced by the
co-accelerated electrons via meson photoproduction and
Bethe-Heitler pair production and, more importantly, with the
strong ambient magnetic field emitting synchrotron radiation
(mesons and muons also emit synchrotron radiation). The SPB model
neglects external photon field components, and this seems
appropriate for BL~Lac objects and their parent population which
possess only weak accretion disks.  \cite{MP2001} have shown that
this model can reproduce the commonly observed double-humped
blazar SED.  Hadronic models require high proton energies that
can only be achieved in a highly magnetized environment where
synchrotron losses can become severe.
Magnetic field values
around $10^3$~G are thought to exist near the horizon of a
supermassive black hole (\cite{BZ77}) with a mass of
$\sim$$10^9$~M$_{\sun}$ as estimated for M87
(\cite{Marconi1997}).  However, with M87's rather low accretion
rate if the equipartition value of $B$ scales with $\dot M$, and
assuming magnetic energy flux conservation, magnetic field
strengths of order 10-100~Gauss are expected within 30
Schwarzschild radii $r_g$ where the jet is probably formed
(\cite{Nature}). 
In the present work, we discuss in more detail
than our earlier work (\cite{Ray2003}), and in the context of the
recent HEGRA detection (\cite{HEGRAdet}), the nuclear (core)
emission, i.e.\ from the M87 jet, in the framework of the
SPB-model.

In Section 2 we summarize the data on M87's core emission.  In
section 3 we give a brief model description, and calculate the
steady-state synchrotron component as described in the appendix.
The modeling procedure is described in section 4, and we conclude
with a summary and discussions in section 5.


\section{The data}
Speculation that M87 could be a powerful accelerator of cosmic
rays triggered space-based $\gamma$-ray detectors and
ground-based high-sensitivity Cherenkov telescopes to search for
$\gamma$-ray emission from this radio galaxy.  Until recently,
only upper limits were obtained.  From EGRET data \cite{EGRET}
obtained $F(>$100~MeV$)<2.2$$\times$$10^{-8}$ cm$^{-2}$ s$^{-1}$.
Using Whipple data from 2000--2001 (\cite{WhippleM87}) and
2002--2003, \cite{Whipple2M87} obtained $F(>$250~GeV$) <2.6\times
10^{-11}$ cm$^{-2}$ s$^{-1}$.  Using 1998-1999 data from HEGRA
telescope array \cite{HEGRAM87} obtained $F(>720$~GeV$)
<1.45$$\times$$10^{-12}$ cm$^{-2}$ s$^{-1}$ (at the $3 \sigma$
level).  By doubling the data set and applying a more sensitive
analysis method, the HEGRA team has recently been published the
first (though tentative) detection of $> 730$ GeV photons at the
$4 \sigma$ level (\cite{HEGRAdet}). This detection, which does
not contradict the Whipple upper limits, places the first data
point in the so-far rather unconstrained high energy regime of
M87's SED, and has motivated us to
refine our previous modeling attempt and predictions for
$\gamma$-rays from this source (\cite{Ray2003}).

The non-simultaneous SED is shown in Fig.~\ref{HBLfig} and
\ref{LBLfig}.  The data imply that the synchrotron spectrum from
the primary electrons must exhibit a break frequency at either a
few $10^{12}$Hz (\cite{Perlman2001}) or around $10^{14}$Hz. Break
frequencies above $10^{14}$Hz as proposed by \cite{BaiLee2001}
seem unreasonable to us given the fact that, firstly, the EUVE
data point from \cite{Berghoefer2000} with its considerably
lower resolution compared to the other data points should
effectively be considered rather as an upper limit, and secondly,
the recent flux measurements from the Chandra observatory
(\cite{WilsonYang2002}) point to a much steeper spectrum in the
X-ray band than anticipated by \cite{BaiLee2001}. Note,
however, that variability effects may play a crucial role
here. Another interesting spectral feature is the strong
steepening by $\Delta \alpha \approx 1$ that occurs around
IR/optical wavelengths, which can not easily be explained by a
transition from escape dominated to synchrotron cooling dominated
electron energy losses.

Images from the VLA (\cite{Biretta1995}), HST (\cite{Sparks1996},
\cite{Biretta1999}), Gemini (\cite{Perlman2001}) and Chandra
(\cite{WilsonYang2002}, \cite{Harris2003}) of the so-far
unresolved core of M87 are at present at the sub-arcsec scale,
giving a metric resolution of order 1-10 pc.  Sub-mas scale cm
and mm wavelength intercontinental VLBI (\cite{Junor1995},
\cite{Nature}) provide the highest linear resolution of
$\sim$0.01 pc achieved on any extragalactic radio jet so far, and
thereby place the most stringent upper limit on the size of the
radio emitting region.  Striking variability in the core region
has not only been observed in the radio to optical band, but has
also been deduced from Chandra X-ray monitoring in 2002
(\cite{Harris2003}). An observed flux increase of about 20\%
which has been measured within 46 days can be transformed into a
doubling time of about 77 days, and provides a limit for the
source size of the X-ray emitting region of $R\simeq 0.1$pc.

HST data (\cite{Biretta1999}) show that features within the first
arcsec of the jet move only at sub-luminal speeds, while at larger
distances from the core super-luminal motion is observed.
In this work we assume that the pattern speed equals
the flow velocity of the knots, or ``plasmoids''.
The upstream knot closest to the core has an apparent speed of
$0.63\pm0.23$c which we use here to constrain the beaming factor
for the nuclear emission.  For a jet angle between
$10\degr-40\degr$ as suggested from VLA and HST proper motion
studies (\cite{Biretta1995}, \cite{Biretta1999}) we find that
Doppler factors in the range $D=1.5$--3 are consistent with the
apparent bulk speed.


\section{The model}

We assume the emission region, or ``blob'', in an AGN jet moves
relativistically with Lorentz factor $\Gamma_j$ and velocity
$\beta_jc$ along the jet axis.  We further assume that
relativistic (accelerated) protons, whose particle density $N'_p$
follows a power-law spectrum $\propto {\gamma_p'}^{-\alpha_p}$ in
the range $2\leq\gamma'_p\leq\gamma'_{\rm{p,max}}$ (primed
quantities are in the jet frame), are injected instantaneously
into a highly magnetized environment ($B'$ is constant within the
emission region), and that they remain quasi-isotropic in the jet
frame due to pitch-angle scattering.  The proton energy-loss
processes considered in the model are photomeson production,
Bethe-Heitler pair production, proton synchrotron radiation and
adiabatic losses due to jet expansion.  Synchrotron radiation
prior to their decay from $\pi^\pm$ (from photomeson production)
and $\mu^\pm$ (from $\pi^\pm$ decay) becomes important in highly
magnetized environments (\cite{RM98}), and is taken into account
in our calculations.  We assume that the maximum particle
energies are limited by the balance between energy gain and loss
rates.  The acceleration rate for any acceleration mechanism is
$dE'/dt' = \xi(E') Ze c B'$ where $\xi(E') \le 1$ is the
acceleration rate factor and $Ze$ is the charge. If particles
gain energy by diffusive shock acceleration and the spectra of
both electrons and protons are cut off by synchrotron losses at
Lorentz factors $\gamma_{e, {\rm max}}$ and $\gamma_{p, {\rm
max}}$, respectively, then
\begin{equation}
{\xi(E_{e, {\rm max}})\over \xi(E_{p, {\rm max}})} = 
\left({m_e \over m_p}\right)^{4(1-\delta)/(1+\delta)}
\end{equation}
if the diffusion coefficient has energy dependence
$\kappa(E)\propto E^\delta$ (see Appendix A).  Thus for a
Kolmogorov spectrum of turbulence, for which $\delta$=1/3,
$\xi(E_{e, {\rm max}})/\xi(E_{p, {\rm max}})=(m_e /
m_p)^2 \approx 3 \times 10^{-7}$.

The relativistic primary $e^{-}$, injected into the emission
region with a power law particle distribution $\propto
E^{-\alpha_e}$, radiate synchrotron photons that manifest
themselves in the blazar SED as the synchrotron hump, and serve
as the target radiation field for proton-photon interactions, and
for the subsequent pair-synchrotron cascade which develops as a
result of photon-photon pair production in the magnetized blob.
The steady-state primary electron spectrum in the co-moving frame
of the emission region is calculated as in Appendix~A taking into
account synchrotron and escape losses. The resulting synchrotron
radiation from this particle distribution is then corrected for
synchrotron-self absorption.  For $B' > 0.6
(u'_{\rm{phot}}/10^{10}\mbox{eV cm}^{-3})^{1/2}$ Gauss the target
photon density $u'_{\rm{phot}}$ is smaller than the magnetic
field energy density. Thus Inverse Compton losses can in most
cases be neglected in the SPB model,
and the corresponding
SSC component is expected to be low
compared to the primary synchrotron 'hump'. However, for
completeness we have calculated explicitely the SSC radiation
from the primary electron component (see Appendix A). 

The pair-synchrotron cascade redistributes the photon power to
lower energies where the photons eventually escape from the
emission region of size $R'$.  The cascades can be initiated by
photons from $\pi^0$-decay (``$\pi^0$ cascade''), electrons from
the $\pi^\pm\to \mu^\pm\to e^\pm$ decay (``$\pi^\pm$ cascade''),
$p$-synchrotron photons (``$p$-synchrotron cascade''), charged
$\mu$-, $\pi$- and $K$-synchrotron photons
(``$\mu^\pm$-synchrotron cascade'') and $e^\pm$ from the
proton-photon Bethe-Heitler pair production (``Bethe-Heitler
cascade'').  Direct proton and muon synchrotron radiation is
mainly responsible for the high energy hump whereas the low
energy hump is dominated by synchrotron radiation from the
primary $e^-$, with a contribution of synchrotron radiation from
secondary electrons produced by the $p$-synchrotron and
$\mu^\pm$-synchrotron cascades.  The contribution from
Bethe-Heitler pair production turned out to be negligible.  For
our calculations we use a Monte-Carlo method and utilize the
recently developed SOPHIA code for the photohadronic event
generation (\cite{SOPHIA}). In practice, to save CPU-time the
target photon field is parametrized as a multiple broken
power-law, which is then used as an input into the SOPHIA code.

\section{Modeling the SED of M87}

From variability and direct imaging arguments an upper limit for
the size of M87's emission region of a few $10^{16}$cm in the
observer frame can be deduced. Together with the constraints from
proper motion measurements, a reasonable parameter space for the
modeling procedure is $R'=10^{15\ldots 16}$cm and bulk Doppler
factors $D=1.5\ldots 3$. Furthermore we demand approximate
equipartition between magnetic $u'_B$ and particle energy density
$(u'_p+u'_e) \approx u'_p$.  This in general minimizes the total
jet power for a given parameter set (\cite{MP2001}).
The observed synchrotron hump in the SED implies a break in the primary
electron spectrum.  A peak in the ``synchrotron
hump'' in the SED is expected at either mm-wavelengths or in the
optical.  In the following we shall model both possibilities in
turn, i.e.\ a high and low energy peak.

\subsection{High-energy peaked synchrotron component (Model H)}

\begin{figure*}
\sidecaption
\includegraphics[width=10.5cm]{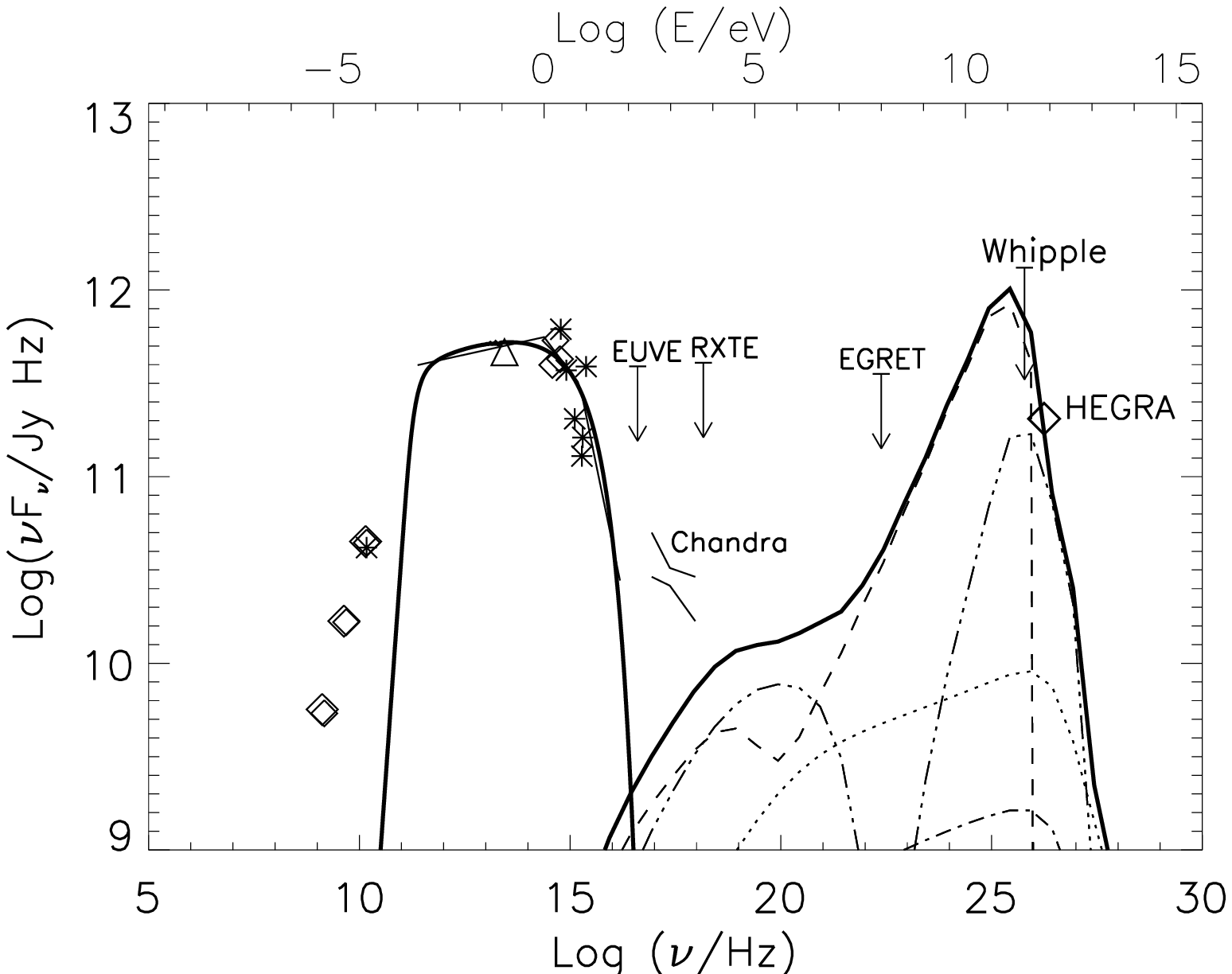}
\caption{ Non-simultaneous SED of M87's core compared with
Model~H fit.  Data are from; \cite{Biretta1991} (diamonds); HST
--\cite{Sparks1996} (stars);
Gemini -- \cite{Perlman2001} (triangle); EUVE --
\cite{Berghoefer2000} (assuming a ratio for the jet/core flux of
1.5), RXTE -- \cite{Reynolds1999}; Chandra --
\cite{WilsonYang2002}; EGRET --\cite{EGRET}; Whipple --
\cite{WhippleM87}; HEGRA -- \cite{HEGRAM87} and \cite{HEGRAdet}.
Flux uncertainties range from $\sim$20\% (radio-to-optical data)
to 26\% (HEGRA data point).  The uncertainty of the Chandra
measurements is indicated. Flux variability may add to these
uncertainties.  Model H parameters are: $B'=30$ G, $D = 2$,
$R'\approx 2$$\times$$10^{15}$ cm, $u'_{\rm{phot}} \approx
3$$\times$$10^{10}$ eV cm$^{-3}$, $u'_p = 15$ erg cm$^{-3}$,
e/p$\approx$ 7.6, $\alpha_e=\alpha_p=1.9$, L$_{\rm{jet}} \approx
2.5$$\times$$10^{43}$erg/s, $\gamma'_{p\rm{,max}} = 3
$$\times$$10^{10}$,
$\xi(E'_{p\rm{,max}}) \approx 1$, $\xi(E'_{e\rm{,max}})=7\cdot 10^{-8}$.
The target photon field for $p-\gamma$
interactions is the primary electron synchrotron photon field,
approximated by broken power laws (thin solid line) with break
energies $\epsilon_{b,1}=1$~eV and $\epsilon_{b,2}=7$~eV between
$8\cdot 10^{-4}$eV and 70~eV in the observer frame and photon
spectral indices $\alpha_1=1.95$, $\alpha_2=2.3$ and
$\alpha_3=3.1$.  The total cascade spectrum (solid line) is the
sum of $p$ synchrotron cascade (dashed line), $\mu$ synchrotron
cascade (dashed-triple dot), $\pi^0$ cascade (dotted line) and
$\pi^{\pm}$-cascade (dashed-dotted line).
The expected SSC
component covers the X-ray regime with a flux level that is not
visible in this figure.
}
\label{HBLfig}
\end{figure*}

The HEGRA-detection at sub-TeV energies places an important
constraint on the models: protons must be accelerated to energies
above $10^{10}$GeV. This can only be achieved by high magnetic
field values, and/or a thin target photon field to prevent
excessive losses at the highest energies. Typical magnetic field
values in the SPB-model lie around several 10 G.
Fig.~\ref{HBLfig} shows a reasonable representation of the data
where we have used $B'=30$ G
($u'_B$=2$\times$$10^{13}$ eV cm$^{-3}$)
and the primary synchrotron component peaking at about
1~eV.  With a Doppler factor of $D = 2$ and an observer-frame
size of the emission region of $R \approx 10^{15}$ cm the target
photon density is $u'_{\rm{phot}}\approx 3$$\times$$10^{10}$ eV
cm$^{-3}$,
and so the SSC component is negligible in Model H.
Proton synchrotron losses dominate at the highest proton
energies for this parameter set (see Fig.~\ref{lossfig}), and 
determine the cutoff energy and the $\gamma$-ray output from
$\sim$ 1 TeV down to $\sim 10$ MeV.  At 0.1-10 MeV synchrotron
radiation from a secondary e$^\pm$ population produced by the
reprocessed $\mu^\pm$/$\pi^\pm$ synchrotron radiation dominates
and produces a broad ``valley'' in the SED between the low and
high energy humps that possess approximately equal power.
An acceleration rate factor at the maximum proton energy
$\xi(E'_{p\rm{,max}}) \approx 1$ is necessary to allow the
injection proton spectrum to extend up to $\gamma'_{p\rm{,max}} =
3 \times 10^{10}$.

The primary electron synchrotron spectrum shows a low-energy
break at the synchrotron self-absorption turnover energy of $\sim
\mbox{a few} \; 10^{-4}$ eV, followed by a nearly flat power
distribution, and then a turnover at about a few eV with a
subsequent steep tail due to the cutoff in the electron
distribution.  Note that interpreting the strong steepening at a
few eV in the data (see Sect.~2) as being due to a cutoff in the
electron spectrum allows spectral breaks larger than 0.5. The
high magnetic field leads to a dominance of synchrotron losses
throughout the emitted low-energy component (the escape loss
dominated energy range lies below the synchrotron-self absorption
turnover frequency).

The total jet power for this parameter choice is L$_{\rm{jet}}
\approx 3$$\times$$10^{43}$erg/s, about the value for the jet
kinetic power derived by \cite{matter},
but still below the nuclear
power of M87 for accretion at the Bondi spherical rate ($\sim 5 \times
10^{44}$ erg/s, \cite{dimatteo}). The small radiative
efficiency of the accretion disk and the fact that the radio power of
M87 is low provides a natural explanation that M87 is a FRI
source. \cite{Owen2000} have calculated the total
bolometric luminosity of the order of $10^{42}$~erg/s, in agreement with
the total radiative output in the present models, which suggests that the jet
in M87 is also a low efficiency radiator and that M87 is currently in
a dormant activity stage.

\subsection{Low-energy peaked synchrotron component (Model L)}

\begin{figure*}
\sidecaption
\includegraphics[width=10.5cm]{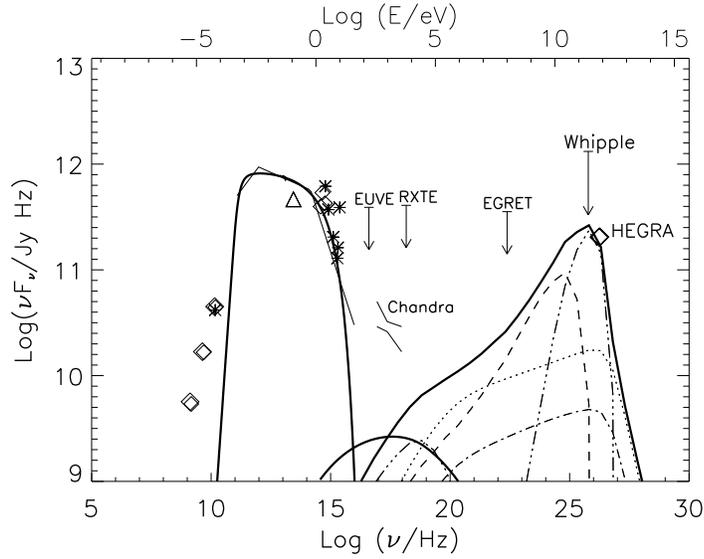}
\caption{ Model L. Parameters are: $B'=5$ G, $D = 1.5$, $R'=
2$$\times$$10^{16}$ cm, $u'_{\rm{phot}} = 2$$\times$$10^{9}$ eV
cm$^{-3}$, $u'_p \approx 1$ erg cm$^{-3}$, e/p$\approx$ 7,
$\alpha_e=\alpha_p=2.1$, L$_{\rm{jet}} \approx 9\times
10^{43}$erg/s, $\gamma'_{p\rm{,max}} = 4$$\times$$10^{10}$,
$\xi(E'_{p\rm{,max}})\approx 1$,
$\xi(E'_{e\rm{,max}})=3\cdot 10^{-8}$.
For the the target
photon field (thin solid line) we used $\alpha_1=1.7$,
$\alpha_2=2.1$, $\alpha_3=2.7$ and break energies
$\epsilon_{b,1}=0.004$~eV and $\epsilon_{b,2}=0.6$~eV between
$2\cdot 10^{-4}$eV and 50~eV in the observer frame.
The expected
SSC emission is shown as the solid curve in the X-ray regime.
}
\label{LBLfig}
\end{figure*}

Fig.~\ref{LBLfig} shows an example for a model in which the primary
electron synchrotron component peaks at $\sim 10^{-3}$eV.  Here
we have chosen the size of the emission region to be of the order
of the limiting size from direct VLBI imaging, i.e.\ $\sim$0.01~pc
(see Sect.~2). With $D = 1.5$ and $R \approx 10^{16}$ cm the
jet-frame target photon density is low, $u'_{\rm{phot}}\approx
2$$\times$$10^{9}$ eV cm$^{-3}$. A relatively low magnetic field
strength of $B'=5$ G gives equipartition between magnetic and
particle energy densities.

\begin{figure*}
\sidecaption
\includegraphics[width=10.5cm]{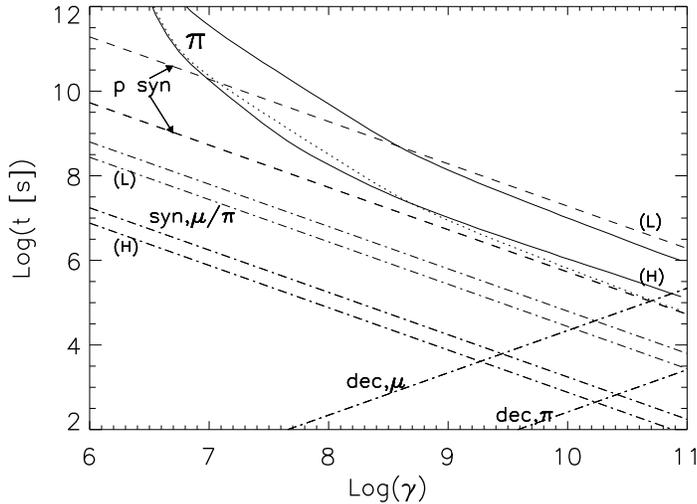}
\caption{ Mean energy loss times (jet frame) for Models (H) and
(L) for $\pi$-photoproduction ($\pi$, solid lines) and proton
synchrotron radiation (p syn, dashed line).  While in model (H) p
synchrotron losses dominate, in model (L) $\pi$ production losses
become dominant at the highest proton energies.  Loss times for
$\pi^\pm$ and $\mu^\pm$ synchrotron radiation (syn $\pi$, syn
$\mu$) are also shown and compared with their mean decay time
scales (dec $\pi$, dec $\mu$). The dotted line represents the
$\pi$ production loss time scale for the same target photon
density as used in model (H) but with a spectral distribution as
in model (L). Lowering the target field's peak energy leads to an
increase of the $\pi$ production rate at the highest proton
energies.}
\label{lossfig}
\end{figure*}

Despite the lower target photon energy density (down by a factor
$\sim$10 compared to model H) pion photoproduction losses
dominate at the highest energies for this parameter set (see
Fig.~\ref{lossfig}) since the magnetic energy density is reduced
by even more (down by a factor 36 compared to model H).  In fact,
even for models with identical magnetic and target photon energy
densities, a lower break-energy in the target spectrum would
cause a turnover in the $\pi$ production losses at
correspondingly higher proton energies and this would in turn
result in a significantly higher $\pi$ production loss rate at
the highest proton energies.  This is demonstrated in
Fig.~\ref{lossfig} where we compare the $\pi$ production loss
rate of the present parameter set (Model L) with the
corresponding rate for the same target spectrum but normalized up
to the photon energy density as used in Model H.  We conclude
that in radiation fields that peak at high energies
$p$-synchrotron losses dominate, while in low-energy-peaked
target photon fields $\pi$ production losses dominate at the
highest proton energies for equal magnetic and target photon
energy densities.

For modeling the primary synchrotron spectrum peaking at around
$10^{-3}$ eV we have injected a softer electron distribution
($\alpha_e=2.1$) into the emission region. The turnover at this
energy is due to synchrotron-self absorption becoming dominant at
radio wavelengths.  A gradual steepening followed by a steep
decline occurs above $\sim$ 1 eV caused by the cutoff in the
electron distribution. Again synchrotron losses dominate the
steady-state electron spectrum above the self absorption turnover
energy.
However, because the ratio $u'_{\rm phot}/u'_B$ is
significantly higher than in Model H, the SSC component (solid
curve at X-ray energies in Fig.~\ref{LBLfig}) is potentially important
in Model L.

Because pion photoproduction losses dominate over proton
synchrotron losses for the present parameter set, the predicted
$\gamma$-ray spectrum above 10 GeV is determined by synchrotron
emission by charged pions and muons rather than by the protons
themselves.  Muon/pion synchrotron radiation naturally produces a
high energy hump extending even up to TeV energies. Because pions
and muons have a lower rest mass than protons, their synchrotron
emission peaks at higher photon energies than the proton
synchrotron radiation for the same Lorentz factor and magnetic
field.  While $\mu^\pm$/$\pi^\pm$ synchrotron radiation dominates
the energy output around the peak energy at $\sim$100 GeV, the
emission in the EGRET energy range is due to proton synchrotron
radiation, and extends down to soft $\gamma$-rays where the
synchrotron radiation from the $\pi$-cascades takes over.  Again,
acceleration rate factor at the maximum proton energy
$\xi(E'_{p\rm{,max}}) \approx 1$ is needed
to explain photon emission up to TeV-energies.

Models with even lower jet-frame target photon densities could
also be consistent with the data (except models with very low
magnetic field values $\leq$ 2 G).  However, the energy density
stored in particles would be orders of magnitude below the
magnetic energy density. If the size of the emission region were
of order $10^{14}$cm or less, the resulting target photon
density would reach values above $10^{11.5}$eV/cm$^3$, and
photopion production losses would cut off the injected proton
spectrum at $\sim 10^9$GeV. In such models TeV-emission at a flux
level as detected by HEGRA would be difficult to explain by
proton or $\mu^\pm$/$\pi^\pm$ synchrotron radiation.

\begin{figure*}
\centering
\resizebox{\hsize}{!}{\includegraphics[width=10.cm]{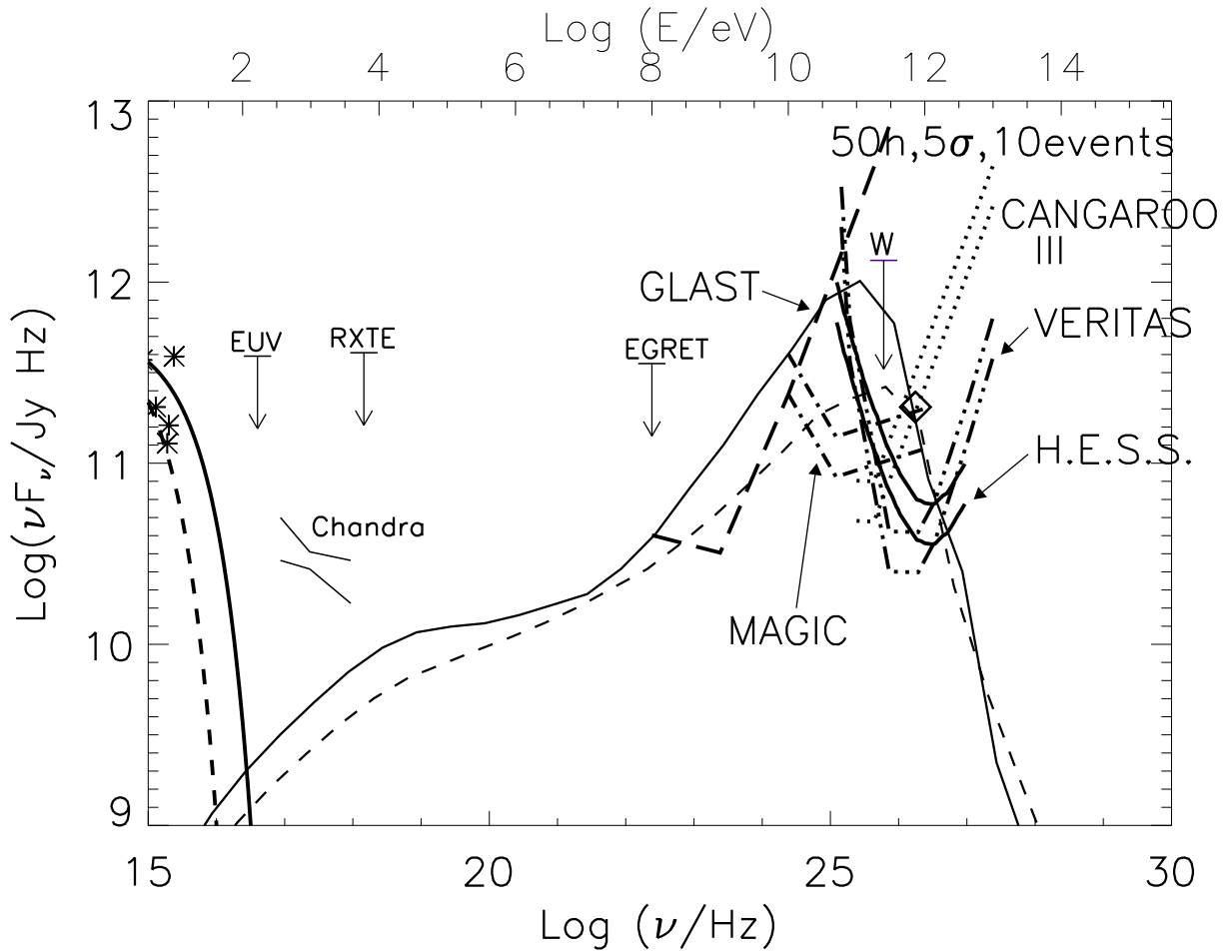}}
\caption{Model H (solid line) and L (dashed line) fits compared
with the sensitivities of $\gamma$-ray telescopes for a source at
zenith and assuming a source photon spectrum $\propto E^{-2.5}$
(thick lower lines) and $\propto E^{-3.5}$ (thick upper lines):
VERITAS: dashed-triple dotted lines; H.E.S.S. phase I: solid
lines; MAGIC (from: ''The MAGIC Telescope Project
Technology and performance aspects'':
http://hegra1.mppmu.mpg.de/MAGICWeb): dashed-dotted lines;
CANGAROO III (from: ''Status Report of the CANGAROO-III
Project'': http://icrhp9.icrr.u-tokyo.ac.jp/c-iii.html): dotted
lines.  For the GLAST sensitivity (from: http://glast.gfsc.nasa.gov/resources/brochures/gsd/)
(long dashed line) a source photon spectrum $\propto E^{-1.5}$ is
used.}
\label{predictfig}
\end{figure*}

\section{Summary and discussion}

We have made SPB model fits to the non-simultaneous SED of M87's
nuclear emission, and find that all parameter sets which
satisfactorily represent the data predict the main contribution
to the high energy luminosity at about 100 GeV to be due to
either $\mu^\pm$/$\pi^\pm$ synchrotron or proton synchrotron
radiation depending on whether the primary electron synchrotron
component peaks at low (Model L) or high (Model H) energies,
respectively.

In the EGRET energy range, the lower synchrotron peak energy
model (Model L) predicts a softer spectrum than the fit with a
higher synchrotron peak energy (Model H).  While it is obvious
that EGRET's sensitivity was more than an order of magnitude
above the expected flux level from M87, we find that the
satellite-based $\gamma$-ray instrument GLAST might possibly
detect a weak signal from this radio galaxy (see
Fig.~\ref{predictfig}).

In all the fits we have presented, the high energy radiation cuts
off with a strong steepening in the TeV range in agreement with
the spectral limits from the HEGRA observation. We also find that
the HEGRA detection at $>$730 GeV can only be explained if the proton
acceleration rate is extremely high ($\xi(E'_{p\rm{,max}}) \approx 1$).
We therefore expect M87, if it is indeed a mis-aligned SPB, could be
an important source of UHECRs (see also \cite{Ray2003}).

For almost all proposed models of particle acceleration in
different astrophysical environments, $\xi(E)$ remains a rather
uncertain model parameter.  On the other hand, any postulation of
acceleration of high energy protons in compact $\gamma$-ray
production regions actually implies that
$\xi(E'_{p\rm{,max}})$ at these energies should be close to
unity, which corresponds to the maximum theoretically possible
acceleration rate based on simple geometrical consideration
(e.g. \cite{Hillas}). An interesting possibility could be
particle acceleration at the annihilation of magnetic fields in
the fronts of poynting flux dominated jets (\cite{Blandford76};
\cite{Lovelace76}).  It has been argued that this mechanism could
provide effective acceleration of extremely energetic protons
with $\xi(E'_{p\rm{,max}}) \sim 1$ (Haswell et al. 1992).  A
quantitative investigation of particle acceleration mechanisms is
beyong the scope of this paper.

Both model fits presented seem to under-predict the emission in
the radio domain as compared to the observations. Our modeling,
however, assumes the same size for the emission region at all
energies, while the data indicate a smaller width of the optical
than the radio jet (\cite{Sparks1996}) though with
{\emph{roughly}} the same morphology.  In addition, the
inter-knot region is observed to be weaker in the optical than in
the radio band (\cite{Sparks1996}). It appears therefore
reasonable to attribute the missing flux in our model to the
inter-knot region, and to a larger blob size in the radio band as
compared to higher frequencies.

As previously noted, the Chandra data lie above the extrapolation
of the optical spectrum to higher energies
(\cite{WilsonYang2002}).
We point out that the modelled SED of
M87 is based on non-simultaneous data, and that the X-ray flux
could have been much lower than the Chandra data suggest at the
time of the optical observations.  Another critical point is that
different resolutions of the images at the various energies have
been used in the literature for the flux determinations.  We have
used for our compilation of the SED the highest resolution data
available at each frequency, ranging from arcmin (HEGRA) to
sub-arcsecond (Chandra, Hubble, VLBI) scales, and could introduce
additional non-negligible uncertainties into the flux
measurements.  However, a flatter X-ray spectrum consistent with
the Chandra data might be achieved if either the magnetic field
is highly inhomogenous, or a secondary e$^\pm$ population is
responsible for the X-ray flux (\cite{WilsonYang2002}).

While, in principle, the SPB model provides secondary synchrotron
emitting e$^\pm$s from the various cascades which may extend even
into the X-ray domain, the fits presented can not easily explain
the high flux level observed at these energies as
electron synchrotron radiation or as part of the high energy hump cascade
component.  Nevertheless, for magnetic fields of order a few
Gauss SSC radiation might become detectable also in hadronic
models.  Fig.~\ref{LBLfig} shows that the SSC component peaks at
X-ray energies with a spectral signature that is in agreement
with the Chandra observations, but its flux is roughly an
order of magnitude too low to explain the Chandra data.  If the
magnetic field or the size of the emission region were a factor
$\sim 3$ lower model L would predict the Chandra X-rays as SSC
photons. The latter change moves model L slightly in the
direction of model H. Thus, X-ray variability might be related to
relatively small changes in the model paramaters, raising or
lowering the importance of inverse Compton losses with respect to
synchrotron losses.

Another possible source for the observed emission in the Chandra
band might be a contribution, either directly or reprocessed
through cascading, from photon fields other than the primary
electrons' synchrotron radiation. The spectral continuum data do
not show any thermal bump from the putative accretion disk.
\cite{dimatteo} have shown that an advection dominated accretion
disk could account for a large fraction of the observed X-ray
nuclear flux. The radiative efficiency is extremely low, but the
accretion rate is found to be large enough for Comptonization of
the synchrotron emission of the disk or the thermal
bremsstrahlung emission to dominate the X-ray emission
(\cite{dimatteo}). A sudden drop of one order of magnitude of the
accretion rate could lower the X-ray disk output by $\sim 2$
orders of magnitude.

For an advection dominated disk with a X-ray luminosity $L_{\rm
disk}\approx 7\times 10^{40}$~erg/s (\cite{dimatteo}) the total
energy density in the jet frame $u'_{\rm disk}$ can be derived
through the transformation (see e.g.\ \cite{DermerSchlicki2002})
\[
 {u'_{\rm disk}} \approx {1.7 \cdot 10^{10}\mbox{eV cm}^{-3}{[\Gamma_j(1-\beta_j} \mu(r))]^{2}}
\left(\frac{L_{\rm disk}}{10^{40}\mbox{erg s$^{-1}$}}\right) \left(\frac{r/\mu(r)}{10^{15}\mbox{cm}}\right)^{-2}
\]
where $r$ is the distance of the jet plasma blob from the black
hole, $\mu(r)=r/(r^2+r_{\rm in}^2)^{1/2}$ and we have
approximated the disk's radiation field as a luminous ring of
radius $r_{\rm in}$ that illuminates the moving blob.  Assuming a
inner accretion disk radius of $r_{\rm in}=1.23 r_g$ (for an
extreme Kerr black hole of mass $M=10^9M_\odot$) and taking
$\Gamma_j=1.5$ one obtains ${u'_{\rm disk}} \sim 2\cdot
10^{10}$eV cm$^{-3}$ at $r=10^{15}$cm and $\sim 2\cdot 10^8$eV
cm$^{-3}$ at $r=10^{16}$cm.  Hence, for $r\ga 10^{16}$cm the
accretion disk radiation proves to be unimportant as a target
field for cascading and photon-particle interactions in M87
compared to the primary electron synchrotron emission.

Recently \cite{DoneaProtheroe2003} have constrained the torus
temperature of the torus to $ <250$~K using existing data from
the literature.  On the other hand, during an extreme flaring
state, related to the accretion rate changes or to a spin flip of
the central black hole, the torus could undergo enough heating to
become 'visible'. This alters the high energy part of the
spectrum above several hundreds of GeV, as is discussed in
\cite{DoneaProtheroe2003}.  Therefore, a visibility-state of the
torus (if present) could be achieved at the cost of not being
able to observe very high energy gamma rays from the nucleus of
M87. Regular monitoring of M87 at VHE gamma-rays and IR
frequencies could be important to elucidate the problem of
existence or non-existence of a dusty torus in M87.  For a
temperature of the torus radiation of $< 250$K the co-moving
frame energy density is $2\cdot 10^{7} \Gamma_j^2$ eV cm$^{-3}
\ll u'_{\rm{phot}}$ for M87, and is therefore negligible as
target photon field. The star and dust contribution of M87's host
galaxy has been estimated to $630 \Gamma_j^2$ eV cm$^{-3}$ and
$6.3 \Gamma_j^2$ eV cm$^{-3}$ in the jet frame, respectively
(\cite{Stawarz2003}), and can obviously also be neglected
regarding M87's synchrotron radiation density of order $10^{10}$
eV cm$^{-3}$.


Fig.~\ref{predictfig} shows that the recently commissioned
Cherenkov telescope array VERITAS, the southern arrays H.E.S.S.\
and CANGAROO III (though at large zenith angles $> 45\degr$),
and MAGIC may be able to detect M87.  The predicted integral
fluxes $> 100$ GeV for Models H and L are $\sim 4\times
10^{-11}$~cm$^{-2}$~s$^{-1}$ and $\sim 4\times
10^{-12}$~cm$^{-2}$~s$^{-1}$, respectively.  We have used
A.~Konopelko's simulator for the H.E.S.S. response
(http://pluto.mpi-hd.mpg.de/~konopelk/WEB/simulator.html) to
estimate the necessary observation time for statistically
significant detection.  A 10~h on-source observation with the
full phase I (four telescopes) H.E.S.S.\ array would result in a
$8-9 \sigma$ detection (expected cosmic ray rate is $\sim$
0.7~s$^{-1}$, $\gamma$-ray rate is 0.055~s$^{-1}$)
in the case of a high-energy peaked photon target, and $4-5
\sigma$ detection in the case of the low-energy peaked photon
target for 300~h of usable data assuming the source at zenith
(expected cosmic ray rate is here $\sim$ 0.7~s$^{-1}$,
$\gamma$-ray rate is 0.006~s$^{-1}$).  Since the sensitivity of
VERITAS (\cite{VERITASsens}) is similar to that of H.E.S.S.,
similar numbers can be expected for VERITAS observations.  In
Fig.~\ref{predictfig} we summarize the minimum flux for a 50~h
observation (with statistics exceeding 10 photons and a signal
detection at a level of at least 5$\sigma$) using the phase I
H.E.S.S.\ array, the VERITAS array, CANGAROO III and MAGIC
(assuming the source at zenith) and GLAST, in comparison the the
predicted high energy fluxes.  Note, however, that these
predictions are based on a non-simultaneous observed SED and,
depending on the actual activity state of M87, the predicted
fluxes and spectra may change significantly. In addition,
absorption of $\gamma$-rays in infrared radiation from a putative
torus could affect the spectrum above 1~TeV if the torus
temperature $T_{\rm{torus}}$ were higher
than 250~K, and above 200~GeV if $T_{\rm{torus}}\geq 1000$~K
(\cite{DoneaProtheroe2003}).
Work is in progress to make SPB model fits to other nearby FR~I
radio galaxies.

In both models presented here, the power output in the high
energy hump is roughly equal to the power output in the low
energy hump of the SED. Because of M87's proximity, absorption of
sub-GeV/TeV-photons in the cosmic infrared background radiation
field is not expected to affect the spectrum below $\sim$ 50~TeV.
{\emph{The observed spectral behaviour at high energies should be
intrinsic to the source.}} Tracing the spectrum at
GeV-TeV-energies would give a $\gamma$-ray spectrum that for the
first time includes an unabsorbed (by radiation fields external
to the source) cutoff.  These data could serve as a typical
template BL~Lac spectrum at source after correcting for M87's jet
mis-alignment. By comparing this template with BL~Lac spectra at
high redshifts, meaningful constraints for the extragalactic
background radiation field around IR wavelengths can be derived.

\begin{acknowledgements}
We thank A. Konopelko for providing us with his H.E.S.S. response
simulator and fruitful discussions. AR's research is funded
by DESY-HS, project 05CH1PCA/6, and that of RJP and ACD
by an ARC Discovery Project grant.
\end{acknowledgements}

\appendix

\section{Steady-state electron spectrum, maximum energies, synchrotron and SSC radiation}

Consider a blob which is moving relativistically with Lorentz
factor $\Gamma_j$ and along the jet axis that is viewed from an
observer at angle $\theta$.  In the jet frame relativistic
electrons are injected into the blob of size R' in the jet
frame. We assume that pitch angle scattering maintains a
quasi-isotropic particle distribution. Our interest is to derive
the steady-state electron spectrum.  In the following all
quantities are in the co-moving frame of the jet, and we omit the
primes for simplicity.  

The number $N(E)dE$ of particles with
energy between $E$ and $E+dE$ in this region is governed by
electron injection, synchrotron cooling process and particle
escape on a time scale $T_{\rm esc}$.  It is described by the
kinetic equation
\begin{equation}
\label{kin}
\frac{\partial}{\partial E}
\left[\dot E N(E) \right] +
   {\frac{N(E)}{T_{\rm esc}}} = Q(E)
\end{equation}
   where
\begin{equation}
\dot E = -\frac{4}{3}\frac{\sigma_{\rm T} c}{m_{\rm e}^2 c^4}
    \left(\frac{B^2} {8 \pi}\right) E^2 = -b E^2\,.
\end{equation}
with $\sigma_{\rm T}= 6.65 \cdot 10^{-25} {\rm cm}^2$ the Thomson
cross section. The first term in Eq.~(\ref{kin}) describes the
rate of energy loss due to synchrotron radiation averaged over
pitch-angle (because of the isotropy of the distribution) in a
magnetic field $B$ (in Gauss), the second term particle escape
from this region at an energy independent rate $T_{\rm
esc}^{-1}=c/R'$. Assuming an injection rate that follows a power
law, $Q(E) = Q_0\cdot E^{-\alpha_e}$, Eq.~(\ref{kin}) has the
solution
\begin{eqnarray}
\label{espec}
N(E) & = & \frac{1}{|\dot{E}(E)|} \int_{E}^\infty \! dE' Q(E')
\exp{\left(-\frac{1}{T_{\rm esc}}\int_{E'}^E \!  \frac{dE''}{\dot
E(E'')}\right)} = \\ \nonumber & = & \frac{Q_0}{b E^2}
\int_{E}^\infty \! dE' E'^{-\alpha_e} \exp{\left[\frac{1}{b
T_{\rm esc}} (E'^{-1}-E^{-1}) \right]} \,.
\end{eqnarray}
For $\alpha_e=2$ and 3 the integral can be solved analytically, giving
\begin{equation}
N(E) = \frac{Q_0 T_{\rm esc}}{E^2} \left[1-\exp{\left(\frac{E_{\rm c}}{E_{\rm max}}-\frac{E_{\rm c}}{E}\right)}\right]
\end{equation}
for $\alpha_e=2$, and
\begin{equation}
N(E) = \frac{Q_0 b T^2_{\rm esc}}{E^2} \left[\frac{E_c}{E}-1-\exp{\left(\frac{E_{\rm c}}{E_{\rm max}}-\frac{E_{\rm c}}{E}\right)
\left(\frac{E_c}{E}-1\right)}\right]
\end{equation}
for $\alpha_e=3$, with $E_c^{-1} = b T_{\rm esc}$ and $E_{\rm
max}$ the maximum injected electron energy. In the case of
injection powers $\alpha_e \neq 2$ or $\neq 3$ we solve the
integral numerically.

The maximum energy, $E_{\rm max}$, is limited by balancing energy
gain and losses. The electrons may gain energy e.g. through
particle acceleration. In general the acceleration time scale may
be written
as $t_{\rm{acc}} = E/[\xi(E) e c B]$ where $\xi(E)
\leq 1$ may be interpreted as an acceleration rate factor.
A comparison with the observations yields typically very low
values $\xi(E_{\rm e,max})$ at the maximum energy for electrons,
while in the case of protons in hadronic models the rate factor
is much higher at the maximum proton energy, typically
$\xi(E_{\rm p,max}) = 10^{-3}...1$.  The large difference
between $\xi(E_{\rm p,max})$ and $\xi(E_{\rm e,max})$ can be
naturally understood from the theory of plasma turbulence, since
the electrons probe much smaller turbulence scales than the
protons (\cite{Rachen2000}).
A more quantitative treatment of
this issue has been presented by \cite{Biermann87}.

In acceleration theory the rate of energy gain is sensitively
dependent on the upstream particle mean free path $\lambda$,
which is given by
\begin{equation}
\lambda(E) =\left.\frac{B^{2}r_{g}}{8\pi I(k)k}\right|_{k=1/r_g}
\end{equation}
in the small angle scattering approximation (\cite{Drury83}) and
with $r_g$ the particle's gyro-radius.  The magnetic turbulence
spectrum $I(k)$ is usually expressed as a power law of the wave
number $k$ in the turbulent magnetic field: $I(k)\propto
k^{-\beta}$.  $\beta=5/3$ corresponds to Kolmogorov turbulence,
while $\beta =1$ corresponds to a fully-tangled magnetic field
resulting in ``Bohm diffusion'', and is often considered for
simplicity.  For strong magnetic fields, Kraichnan turbulence
$\beta =3/2$ may be present. The (parallel) diffusion coefficient
is then given by $\kappa_{||}=\frac{1}{3}\lambda_{||}v$ where $v$
is the particle's speed and $\lambda_{||}$ is its mean free path
parallel to the magnetic field.  Hence, $\kappa_{||} \propto
E^\delta$ where $\delta=(2-\beta)$, and the acceleration time
scale for the relativistic electrons and protons ($r_g \propto
E$) can then be expressed by $t_{\rm acc}\propto E^\delta \propto
E/\xi(E)$. In the following, we consider $\delta$ to be a free
parameter, and restrict our considerations to parallel shock
fronts only for simplicity. Obviously,
$\xi_p(E)=\xi_e(E)=\xi(E)\propto E^{1-\delta}$ applies to both,
electrons and protons. If the electron and proton spectra are
limited by synchrotron losses, \cite{Biermann87} found for their
cutoff energy
\begin{equation}
{\gamma_{\rm e,max} \over \gamma_{\rm p,max}} \propto
\left(m_e/m_p\right)^{(3-\delta)/(1+\delta)}, \;\;
{E_{\rm e,max} \over E_{\rm p,max}} \propto
\left(m_e/m_p\right)^{4/(1+\delta)}
\end{equation}
and so one expects 
\begin{equation}
{\xi(E_{\rm e,max}) \over \xi(E_{\rm p,max})} \propto
\left(m_e/m_p\right)^{4(1-\delta)/(1+\delta)}.
\end{equation}
The ratio of their maximum synchrotron photon
energies can readily be computed to:
\begin{equation}
\frac{\epsilon_{\rm syn,e}}{\epsilon_{\rm syn,p}} =
\left(\frac{m_e}{m_p}\right)^{(5-3\delta)/(1+\delta)}\,.
\end{equation}
The acceleration model parameters used to calculate the 
SEDs of M87 can be understood for $\delta \sim 0.3$ which is close
to that for a Kolmogorov turbulence spectrum.

To obtain the synchrotron specific luminosity, for the case of no
synchrotron self-absorption, as a function of frequency $\nu$ we
convolve the particle density N(E) with the synchrotron Green's
function $P(\nu,E)$:
\begin{equation}
\label{synemi}
L_0(\nu)\,=\,\int dE \, P(\nu,E) \, N(E)
\end{equation}
with
\begin{equation}
\label{synpower}
P(\nu,E) \,=\, \frac{\sqrt{2}}{2\pi c} e^2 \omega_e
F\left(\frac{\nu}{\nu_c}\right) \,.
\end{equation}
for relativistic particles of velocity $\approx c$ and an
isotropically distributed magnetic field. $\omega_e = e B/m_e$ is
the electron gyro frequency and $\nu_c = \sqrt{\frac{3}{2}}
\omega_e E^2/(m_e^2 c^5)$ the critical frequency after pitch
angle averaging. The function $F(x)$ with $x=\nu/\nu_c$ can be
approximated by (Melrose \cite{melrose80})
\begin{equation}
F(x) = x \int_{x}^\infty \! dx' \, K_{5/3}(x') \approx 1.85
x^{1/3} \exp(-x) \,.
\end{equation}

Synchrotron self-absorption will dominate the photon spectrum at
low energies. The synchrotron radiation and its corresponding
electron spectrum may be approximated by a multiple broken power
law.  For each part of the particle spectrum that is governed by
a simple power law with index $\alpha_{e,i}$ one can therefore
use the absorption coefficient for synchrotron radiation of a
power law electron distribution in a randomly oriented magnetic
field (Longair \cite{longair2}), which reads
\begin{equation}
\chi_{\rm \nu,ssa} = \frac{\sqrt{2} e^3}{8\pi m_e}
\left(\frac{\sqrt{6} \, \omega_e}{2\pi m_e^2
c^5}\right)^{\alpha_e/2} q_0 \, B \,
\Gamma\left(\frac{3\alpha_{e,i}+2}{12}\right) \,
\Gamma\left(\frac{3\alpha_{e,i}+22}{12}\right)
\nu^{-(\alpha_{e,i}+4)/2} \,
\end{equation}
where $q_0$ is $Q_0$ divided by the source volume.
The  synchrotron specific luminosity  in the jet frame is then
\begin{equation}
L(\nu)\,=\, \frac{L_0(\nu)}{\tau_{\rm \nu,ssa}}
\left[1-\exp(-\tau_{\rm \nu,ssa})\right] \,.
\end{equation}
with $\tau_{\rm \nu,ssa} = R \, \chi_{\rm \nu,ssa}$.  Note, that
for high magnetic fields that are typical for the SPB-model, the
electron spectrum, and thus its corresponding synchrotron
spectrum above the synchrotron-self absorption break energy, is
often completely determined by synchrotron losses for typical
blazar 'blob' sizes.

This synchrotron component represents the target photon field for
photopion production and cascading in the SPB-model, and
simultaneously manifests itself as the 'synchrotron hump' in the
blazar SED after transformation of the luminosity $L(\nu)$ into
the observer frame.  To save CPU-time we fit this target photon
field with a multiple broken power law, which is then used as an
input into the SPB Monte-Carlo code (see Sect.~3).

The importance of Inverse Compton scattering off the
synchrotron photons produced by the same primary electron
component is determined by the ratio of the synchrotron photon
energy density and the magnetic field energy density. For $B >
0.6 (u_{\rm{phot}}/10^{10}\mbox{eV cm}^{-3})^{1/2}$ Gauss the
target photon density $u'_{\rm{phot}}$ is smaller than the
magnetic field energy density, which is typically true for
hadronic SPB models, and inverse Compton losses of the primary
electron population can usually be neglected.  In the following we
calculate the expected SSC (jet frame) specific luminosity from an
electron spectrum Eq.~(\ref{espec}) in the Thomson limit which is
valid for $E \ll (m_e c^2)^2\, / h\nu$.  In the $\delta$-function
approximation it is given by:
\begin{equation}
L_{SSC}(E_\gamma) = c \sigma_T E_\gamma \int_0^{\infty} d\epsilon\, n(\epsilon)
\int_{\gamma_{\rm e,min}}^{(m_e c^2)/\epsilon} dE\, N(E)\, \delta\left(E_\gamma-\gamma_e^2\epsilon\right)\,. \nonumber
\end{equation}
where $\gamma_{\rm e,min}m_ec^2$ is the minimum injected electron
energy and $n(\epsilon)$ is the (jet frame) synchrotron photon
density.  For the calculations in Fig.~\ref{HBLfig} and
\ref{LBLfig} we performed the $\epsilon$-integration numerically.

\end{document}